\documentstyle[epsfig]{mn}

\title[Variability in the IMF]{Variability in the stellar initial mass
function at high mass: coalescence models for starburst clusters}

\author[Mohsen Shadmehri]
{Mohsen
Shadmehri\thanks{E-mail: mshadmehri@science1.um.ac.ir}\\
Department of Physics, School of Science, Ferdowsi University,
Mashhad, Iran}

\date{Received / Accepted  }

\begin{document}

\maketitle

\label{firstpage}

\markboth{Shadmehri: Variability in the IMF}{}

\begin{abstract}
A coalescence model using the observed properties of pre-stellar
condensations (PSCs) shows how an initially steep IMF that might
be characteristic of primordial cloud fragmentation can change
into a Salpeter IMF or shallower IMF in a cluster of normal
density after one dynamical time, even if the PSCs are collapsing
on their own dynamical time. The model suggests that top-heavy
IMFs in some starburst clusters originate with PSC coalescence.
\end{abstract}


\section{Introduction}

A recent study of observations of the stellar initial mass
function (IMF) suggest there are systematic variations where the
IMF gets flatter, or more top-heavy, in denser regions (Elmegreen
2004). This paper proposes that the high mass part of the IMF
varies with density as a result of the coalescence of dense
pre-stellar condensations (PSCs), such as those observed by
Motte, Andr\'e, \& Neri (1998), Testi \& Sargent (1998), Onishi
et al. (2002) and Nomura \& Miller (2004). These objects have
densities in the range from $10^5$ cm$^{-3}$ to $10^7$ cm$^{-3}$,
and masses from 0.1 M$_\odot$ to a few M$_\odot$, giving them
sizes of $\sim 10^4$ AU. The largest PSCs may be self-gravitating
(Johnstone et al. 2000, 2001; Motte et al. 2001; Umemoto et al.
2002), as may those with stars (Tachihara et al. 2002). The size
and spatial proximity of PSCs in a cloud core suggest their
coalescence might be important for the highest masses (Elmegreen
\& Shadmehri 2003). This process is modelled to illustrate its
possible effect on the IMF.

\section{The model}

Consider the evolution of a distribution $n(r,M,t)$ of PSCs with
masses $M$ and radii $R_P$ at positions $r$ inside a cloud having
a mass $M_{\rm c}$ and radius $R_{\rm c}$. The cloud density
profile is
\begin{equation}
\rho_c(r)={{\rho_{\rm c0}}\over{1+(r/R_{\rm c0})^2}}
\end{equation}
where the core radius is $R_{\rm c0}$ and the central density is
given by the cloud mass:
\begin{equation}
\rho_{\rm c0}= \frac{M_{\rm c}}{4\pi R_{\rm c0}^3 [R_{\rm
c}/R_{\rm c0}- \arctan (R_{\rm c}/R_{\rm c0})]}.
\end{equation}

The PSCs are taken to have density profiles (Whitworth \&
Ward-Thompson 2001)
\begin{equation}
\rho_{\rm p}\left(r_p\right)={{\rho_{\rm po}}\over{[1+(r_p/R_{\rm
p0})^2]^2}}
\end{equation}
where central density $\rho_{\rm p0}$ is assumed to be constant
and the radius $R_{\rm p}$ varies with condensation mass $M_p$ and
position $r$ in the cloud.  The positional variation assumes the
density at the edge of the condensation equals the ambient cloud
density, $\rho_{\rm p}\left(R_p\right)=\rho(r)$, so PSCs of the
same mass are smaller near the cloud core. Thus the density
contrast between the condensation center and the edge is
\begin{equation}
{\cal C}(r)=\left(1+\left[r/R_{\rm c0}\right]^2\right)\rho_{\rm
p0}/\rho_{\rm c0},
\end{equation}
and the PSC radius can be written as
\begin{equation}
R_{\rm p}(r, M)= a(r)R_{\rm p0}\end{equation} where
\begin{equation}
R_{\rm p0}=(M/2\pi \rho_{\rm
p0})^{1/3}\left(\arctan\left[a(r)\right]-\frac{a(r)}{1+a(r)^2}\right)^{-1/3}.
\end{equation}
and
\begin{equation}
a(r)=\sqrt{\sqrt{{\cal C} (r)}-1};
\end{equation}
$R_{\rm p}\left(r,M\right)$ is assumed to be the PSC radius at the
beginning of its collapse.

The average radius of the PSC used in the collision rate should in
general be smaller than $R_{\rm p}$ because of PSC collapse. Each
PSC is assumed to collapse at a rate proportional to its central
dynamical rate, $\left(G\rho_{\rm p0}\right)^{1/2}$, so each PSC
radius drops effectively to zero as $e^{-\nu\left(G\rho_{\rm
p0}\right)^{1/2}t}$ for parameter $\nu\sim1$.  For a distribution
of formation times $\tau$ up to the current time $t$, these
assumptions give the average PSC radius for use in the collision
cross section:
\begin{eqnarray}
<R_{\rm p}(r, M,
t)>=\qquad\qquad\qquad\qquad\qquad\qquad\qquad\nonumber
\\\qquad \frac{\int_{0}^t R_{\rm p}(r, M) n(r, M, \tau)
\exp(-\nu\sqrt{G\rho_{\rm p0}}\left[t-\tau\right])
d\tau}{\int_{0}^t n(r, M, \tau) d\tau}&.\label{eq:radius}
\end{eqnarray}

Now that the PSC radius is defined, the PSC collision rate and
the evolution of the IMF can be calculated. Two components are
assumed for the IMF at intermediate and high mass. One is
intrinsic to the formation of PSCs and independent of
coalescence. This could be a power law if the PSC formation
process is scale-free, or it could be log-normal if there is a
characteristic mass and some type of random multiplicative
fragmentation or build-up.  To illustrate how apparently power
law IMFs can result from IMFs with a characteristic mass, the
model considers a log-normal intrinsic spectrum of PSC mass,
$F(\log M)d\log M=f(M)dM$ \footnote{In this paper, $\log$ refers
to base 10 }(to use the notation of Scalo 1986). This is taken to
be normalized so that $\int_{M_{\rm min}}^{M_{\rm max}} f(M)dM=1$
for minimum and maximum stellar masses $M_{\rm min}$ and $M_{\rm
max}$. Power law models were also made and the results were
qualitatively the same as those reported here so they are not
discussed further.

The time-evolution of the number, $n$, of stars between masses $M$
and $M+dM$ at time $t$ and distance $r$ from the center of the
cloud is
\begin{eqnarray}
\frac{\partial n(r, M, t)}{\partial t} =  {{\epsilon\rho_{\rm
c}(r)\left(G \rho_{\rm c}[r]\right)^{1/2}}\over{<M>}}
f(M)\qquad\qquad\qquad\qquad\nonumber\\
+ 0.5 \eta \int_{M_{\rm min}}^{M-M_{\rm min}} n(r, \xi, t) n(r,
M-\xi, t) \sigma (\xi, M-\xi) v d\xi\nonumber \\ - \eta
n(r,M,t)\int_{M{\rm min}}^{M_{\rm max}} n(r,\xi,t) \sigma (\xi, M)
v d\xi,\label{eq:main}
\end{eqnarray}
where $<M>$ is the average mass in the IMF, $\epsilon$ and $\eta$
are efficiencies, and $\sigma$ and $v$ are the coalescence cross
section and relative velocity.

The first term is the spontaneous star formation rate. It is
written to give a local rate $\epsilon\rho\left(G\rho_{\rm
c}[r]\right)^{1/2}$ in units of mass per volume per time, where
$\epsilon$ is the fraction of the gas mass converted into star
mass in each unit of dynamical time, $\left(G\rho_{\rm
c}\right)^{-1/2}$ (see Elmegreen 2002). Typically
$\epsilon\sim0.2-0.5$ from observations of cluster formation
efficiencies and timescales (Lada \& Lada 2003; Elmegreen 2000).
When divided by the average star mass in the IMF, $<M>=\int
Mf(M)dM$, it gives the number of stars formed per unit volume and
time. The function $f(M)dM$ is the probability that a star has the
mass $M$ between $M$ and $M+dM$.  Multiplication of the count rate
by f(M) gives the number of stars forming per unit volume and time
in a $dM$ mass interval around $M$. Note that the star formation
mass per volume and time is $\int Mn(r,M,t)dM$, which is
$\epsilon\rho\left(G\rho_{\rm c}\right)^{1/2}$ again, since $\int
Mf(M)dM=<M>$ cancels the $<M>$ in the denominator of equation
(\ref{eq:main}).

In the composite IMF model, $f(M)$ should have a steep slope at
intermediate to high mass so the overall IMF has a steep slope
($|\Gamma|\geq1.7$) in low density environments, as is apparently
observed (Elmegreen 2004). The Miller-Scalo log-normal function
is used here because it has this property. Scalo (1986) cautioned
that the Miller-Scalo function is not a good representation of the
field-star IMF. He replaced the log-normal with a slightly steeper
function at intermediate mass and a slightly shallower function at
high mass. Here, the Miller-Scalo log-normal is used only as a
starting point for the calculation, calling it a spontaneous IMF,
and then it is evolved with coalescence to get a final IMF. For
this log-normal, which is defined for equal intervals of
$\log(M)$, the probability function $f(M),$ which is defined for
equal intervals of $M$, is given by:
\begin{equation}
f(M)={{f_{\rm 0}e^{-B\left(\log\left[M/M_0\right]\right)^2}} \over
{M}},
\end{equation}
where $f_{\rm 0}$ is the normalization factor making $\int
f(M)dM=1$.

The second term in equation (\ref{eq:main}) represents the
collision rate to form a PSC of mass M per unit intervals of
time, volume, and mass.  The cross section $\sigma$ includes
gravitational focusing,
\begin{equation}
\sigma\left(M_x,M_y\right)=\pi\left(R_{px}+R_{py}\right)^2
\left[1+{{2G\left(M_x+M_y\right)}\over{2v^2\left(R_{px}+R_{py}\right
)}}\right].
\end{equation}
The collision velocity $v$ for PSCs is not well observed. It may
be the same for all objects and comparable to the virial velocity
in the whole cloud, or may be a function of distance from the
center, $v\propto r^\alpha$, in analogy with some turbulence
models. Here, $v$ is assumed to equal a representative
three-dimensional virial velocity,
\begin{equation}
v=(3GM_{\rm c}/5R_{\rm c})^{1/2}.
\end{equation}
The coefficient in front of this expression to correct for the
non-uniform mass distribution in the cloud is relatively
unimportant; most of its effect can be absorbed in the factor
$\eta$. The third term in equation (\ref{eq:main}) is the rate of
loss of condensations of mass $M$ resulting from collisions with
all other masses (see Field \& Saslaw 1965).

Equation (\ref{eq:main}) is solved as an initial value problem by
assuming $n(r, M, 0)=0$ at $t=0$. The number density of PSCs
between masses $M$ and $M+dM$ starts out like $f(M)$ for all $r$,
but once a few PSCs form, collisions convert low and intermediate
mass stars into high mass stars, changing the IMF. Because most of
the cloud mass is at low density where collisions are relatively
unimportant, the IMF is not affected much overall, but at small
radii it can change a lot, becoming flatter at high mass.

\begin{figure}
\epsfig{figure=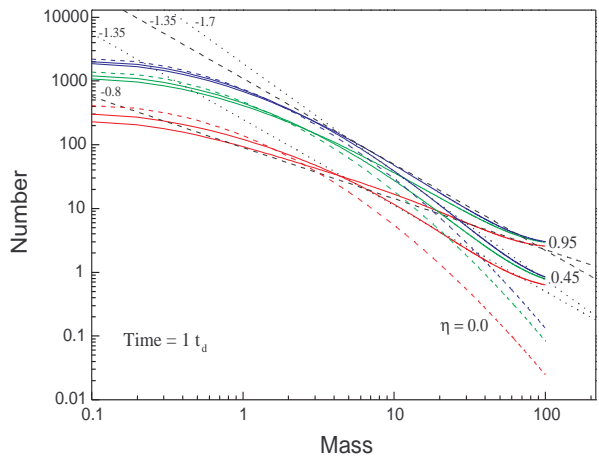,angle=0,width=\hsize} \caption{IMF model
with collisions between pre-stellar condensations following the
evolution equation (14) up to one dynamical time in the cloud
core. Curves are labelled by the value of $\eta$ (dashed curves
have no coalescence: $\eta=0$). Color indicates radius over which
the IMF is integrated: red, green and blue integrate out to 2, 5,
and 10 times the cloud core radius. The IMF is shallower closer
to the center of the cloud because coalescence is more important
there. Fiducial slopes are shown by straight dashed lines with the
$\Gamma$ values indicated. The Salpeter function has
$\Gamma=-1.35$. The IMF at intermediate-to-high mass is steep in
the absence of collisions ($\Gamma\sim-1.7$), as might apply in
low density star-forming regions in the field, and it becomes
shallower over time, especially in the cloud core, as coalescence
makes more massive stars.}\label{fig:model}
\end{figure}

Figure \ref{fig:model} shows several solutions to equation
(\ref{eq:main}).  The plotted quantity is an integral over radius
in the cloud, $2.3M\int_0^R n(r,M,t)4\pi r^2 dr$, where the
multiplication by $M/\log(e)=2.3M$ converts the function $n$ to
$\log$ intervals for plotting. The cloud parameters are:
$M_c=10^4$ M$_\odot$, $R_{\rm c0}=0.1$ pc, and $R_{\rm c}= 5$ pc.
The PSC central density is $\rho_{\rm p0}=10^8\mu$ cm$^{-3}$ for
mean molecular weight $\mu=4\times10^{-24}$ gm. The collapse rate
parameter for PSCs is $\nu=1$. The efficiencies in the evolution
equation are: $\epsilon=0.5$ and $\eta=0,$ $0.45$ and 0.95. Larger
$\eta$ increases the effect of collisions and flattens the IMF;
$\eta=0$ shows the spontaneous mode.  The intrinsic IMF has
parameters as in the other examples above: $M_{\rm min}=M_0=0.1$
M$_\odot$ and $B=1.08$.  The red, green and blue lines represent
IMFs integrated out to 2, 5, and 10 times the cloud core radius,
$R_{\rm c0}$; they illustrate how the cloud core IMF is shallower
than the others.  The total integration time is one dynamical
time in the cloud core, $t_{\rm d}=\left(G\rho_{\rm
c0}\right)^{-1/2}$.

Evidently coalescence in dense clusters can turn a steep intrinsic
IMF into one that resembles a Salpeter function, with a slope of
$\sim-1.35$ or shallower. The same spontaneous IMF without
collisions would have a slope of $\-1.7$ or steeper, depending on
the mass range.

\section{Summary}

We considered a cluster with nonuniform density profile and known
mass and radius, in which PSCs can move approximately with virial
velocity. Considering collisions between PSCs and their
gravitational collapse, a differential equation is presented and
solved numerically which gives number density distribution of
PSCs inside the cloud at each radius, mass interval and time. In
order to control the collision and the collapse rates in a
phenomenological way, some input parameters are introduced. We
explored a wide range of input parameters and the results were
qualitatively the same as those reported here. In some starburst
regions a high fraction of high mass stars compared to low mass
stars has been reported which means a high mass bias (e.g.,
Sternberg 1998; Smith \& Gallagher 2001). Our results suggest
this shift in high mass part of IMFs in these starburst clusters
may originates from PSCs coalescence.

Elmegreen (2004) discussed how various physical processes
operating in at least three mass regimes  may lead to  three
components IMF. This point is demonstrated by reviewing the
observations of IMF variations and using two simple models
showing the observed power law distributions do not necessarily
imply a single scale free star formation mechanism (Elmegreen
2004). Our work in this paper confirms this expectation by
presenting a simple model for collisions between PSC in clusters
and its possible effect on IMF. Of course, we did not intend to
fit the IMF with a particular coalescence model, but showing that
the high mass part of IMF in a composite model as suggested by
Elmegreen (2004), may change due to interactions between PCS.

{}
\end{document}